\documentclass[twocolumn]{aastex631}
\usepackage{amsmath}	
\usepackage{amssymb}	

\shorttitle{Pseudostreamer CMEs}
\shortauthors{Wyper et al.}
\graphicspath{{./}{}}

\usepackage{color}

\newcommand{\jgra}{JGRA}

\newcommand{\grad}{\boldsymbol{\nabla}}
\newcommand{\vb}{\boldsymbol{v}}
\newcommand{\Bb}{\boldsymbol{B}}
\newcommand{\gb}{\boldsymbol{g}}
\newcommand{\evr}{\boldsymbol{e}_r}

\begin{document}

\title{A Model for Flux Rope Formation and Disconnection in Pseudostreamer Coronal Mass Ejections}

\author[0000-0002-6442-7818]{P.~F.~Wyper}
\affiliation{Department of Mathematical Sciences, Durham University, Durham, DH1 3LE, UK}
\email{peter.f.wyper@durham.ac.uk}

\author[0000-0001-6886-855X]{B.~J.~Lynch}
\affiliation{Department of Earth, Planetary, and Space Sciences, University of California--Los Angeles, Los Angeles, CA 90056, USA}

\author[0000-0002-4668-591X]{C.~R.~DeVore} 
\affil{Heliophysics Science Division, NASA Goddard Space Flight Center, 8800 Greenbelt Rd, Greenbelt, MD 20771}

\author[0000-0001-6289-7341]{P.~Kumar} 
\affil{Heliophysics Science Division, NASA Goddard Space Flight Center, 8800 Greenbelt Rd, Greenbelt, MD 20771}

\author[0000-0003-0176-4312]{S.~K.~Antiochos} 
\affil{Department of Climate and Space Sciences and Engineering, University of Michigan, Ann Arbor, MI 48109, USA}

\author[0000-0002-1198-5138]{L.~K.~S.~Daldorff} 
\affil{The Catholic University of America, 620 Michigan Ave., N.E. Washington, DC 20064 USA}
\affil{Heliophysics Science Division, NASA Goddard Space Flight Center, 8800 Greenbelt Rd, Greenbelt, MD 20771}



\begin{abstract}
Coronal mass ejections (CMEs) from pseudostreamers represent a significant fraction of large-scale eruptions from the Sun. In some cases, these CMEs take a narrow jet-like form reminiscent of coronal jets; in others, they have a much broader fan-shaped morphology like CMEs from helmet streamers. We present results from a magnetohydrodynamic simulation of a broad pseudostreamer CME. The early evolution of the eruption is initiated through a combination of breakout interchange reconnection at the overlying null point and ideal instability of the flux rope that forms within the pseudostreamer. This stage is characterised by a rolling motion and deflection of the flux rope toward the breakout current layer. The stretching out of the strapping field forms a flare current sheet below the flux rope; reconnection onset there forms low-lying flare arcade loops and the two-ribbon flare footprint. Once the CME flux rope breaches the rising breakout current layer, interchange reconnection with the external open field disconnects one leg from the Sun. This induces a whip-like rotation of the flux rope, generating the unstructured fan shape characteristic of pseudostreamer CMEs. Interchange reconnection behind the CME releases torsional Alfv\'{e}n waves and bursty dense outflows into the solar wind. Our results demonstrate that pseudostreamer CMEs follow the same overall magnetic evolution as coronal jets, although they present different morphologies of their ejecta. We conclude that pseudostreamer CMEs should be considered a class of eruptions that are distinct from helmet-streamer CMEs, in agreement with previous observational studies.  
\end{abstract}

\keywords{Sun: corona; Sun: magnetic fields}


\section{Introduction} \label{sec:intro}
Coronal mass ejections (CMEs) exhibit a variety of morphologies in coronograph images. In recent years it has become increasingly recognised that these different morphologies are closely tied to the large-scale structures that define the open-closed magnetic boundary overlying the source regions of the events. The structures are of two types: helmet streamers and pseudostreamers. Helmet streamers lie between coronal holes of opposite magnetic polarity and taper into the base of the heliospheric current sheet. Pseudostreamers by contrast lie between or within coronal holes of a single polarity, are associated with at least one coronal null point, and have no large-scale current sheet \citep{Titov2012,Kumar2021}. 

Helmet-streamer CMEs are well studied as most active regions lie beneath helmet streamers. Large active-region or quiet-Sun filament eruptions generally create classic 3-part CMEs with bubble-like shapes \citep[e.g.,][]{Riley2008,Webb2012}. Streamer-blowout CMEs, which form from the large-scale expansion and pinch-off of 
{magnetic loops from the streamer 
and sometimes are associated with filament ejections from the streamer base}
\citep{Vourlidas2018}, are slower but share a similar bubble-like morphology. Many ``stealth'' CMEs \citep{Lynch2016,Bhowmik2022} also fall into this category, as they are the extreme end of the continual pinch-off and blob-formation process occurring at the tips of helmet streamers \citep[e.g.,][]{Sheeley1999,Higginson2017}. Helmet-streamer CMEs therefore all share a generally bubble-like morphology, although they vary considerably in the amount of magnetic flux and plasma ejected. 

\begin{figure}
    \includegraphics[width=0.5\textwidth]{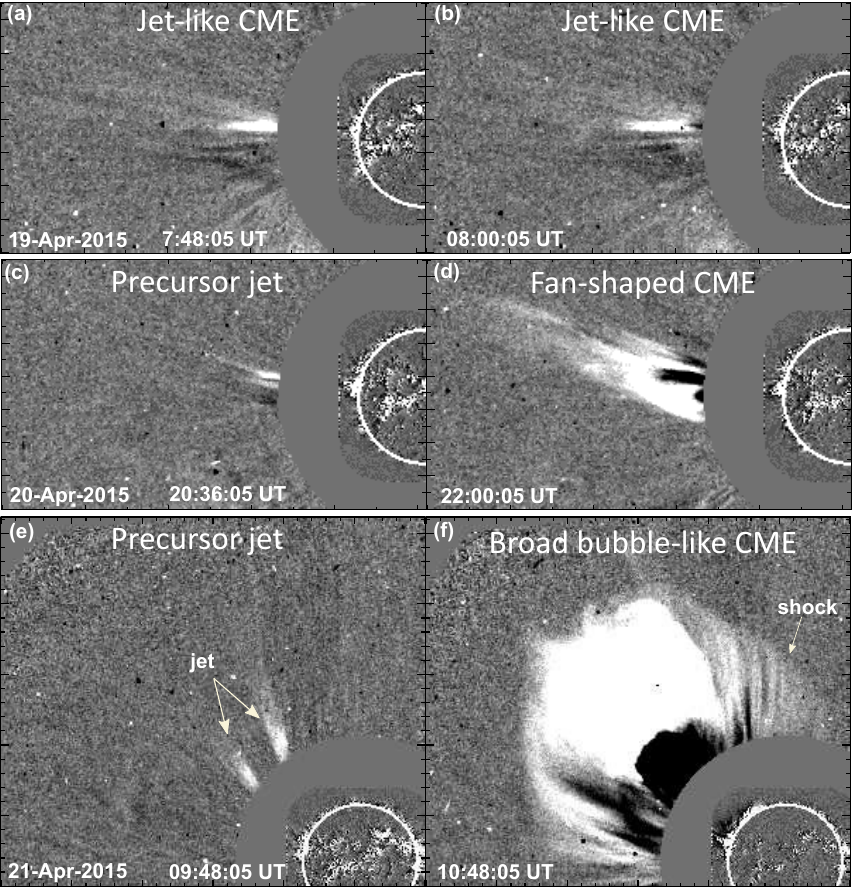}
    \caption{{LASCO C2 running difference images \citep{Brueckner1995} showing the different morphologies of CMEs originating from pseudostreamers. (a) \& (b): two narrow jet-like events. (c) \& (d) a fan-shaped CME and its precursor jet. (e) \& (f) a broad bubble-like CME and its precursor jet. Adapted from \citet{Kumar2021}.}}
    \label{fig:obs}
\end{figure}

CMEs originating from pseudostreamers, in contrast, are more varied in morphology. \citet{Wang2015} and \citet{Wang2018} classified them as either fan-shaped or jet-like. Fan-shaped CMEs have an unstructured core with typical widths up to $30^\circ$, while jet-like CMEs are more collimated with narrower widths nearer $10^\circ$. {Examples of each type are shown in Figure \ref{fig:obs}(a)-(d)}. Both types generally travel at a steady ejection speed once underway. Some fan-shaped CMEs exhibit a V-shape suggestive of concave-up field lines beneath the flux rope. \citet{Wang2023} compared a variety of pseudostreamer CMEs, concluding that all are laterally confined by the adjacent open field and are likely different manifestations of large-scale coronal jets. In contrast, \citet{Kumar2021} analyzed three pseudostreamer CMEs that did not fit the Wang \& Hess pattern. Their more energetic events were much wider than $40^\circ$, were more bubble-like in morphology, and clearly had a shock front ahead of them, {Fig. \ref{fig:obs}(f)}. These characteristics are much more typical of helmet-streamer CMEs. In addition, Kumar et al.\ observed pre-eruption jets {(Fig. \ref{fig:obs}(c) \& (e))} and dimmings associated with slow reconnection/opening near the null points about 1-3 hr prior to the filament/flux-rope eruptions.

All pseudostreamer eruptions involve certain fundamental constituent parts. Filament channels (with or without filament material) form slowly over the course of days to weeks under the arcades of the pseudostreamer before becoming unstable and erupting. These filament channels appear as cavities when viewed on the limb \citep[e.g.,][]{Guennou2016,Karna2019} and can contain either a flux rope or sheared arcade. In the latter case, a flux rope will form once flare reconnection commences beneath the eruption. The novel constituent part of a pseudostreamer (compared to a helmet streamer) is its overlying coronal magnetic null point(s) and background unipolar open flux. Clearly, the subsequent interaction of the rising flux rope with the magnetic topology could be the key determinant of the different pseudostreamer CME morphologies.

Jet-like CMEs are the most straightforward to explain. In this case there is relatively little expansion of the flux rope, which launches a jet-like CME when it reconnects at the pseudostreamer null point. Exactly the same evolution occurs in coronal jets associated with mini-filament eruptions, which share a similar null-point structure but on a much smaller scale \citep[e.g.,][]{Sterling2015,Kumar2018,Kumar2019b,Kumar2019a}. This correspondence suggests that all these events are part of a continuum of jet-like eruptions \citep{Wyper2017,Wyper2021,Kumar2021}. 

In \citet{Wyper2017}, we presented a model for coronal jets with mini-filaments. The model generalises the breakout mechanism for CMEs \citep{Antiochos1999,Lynch2008} to the null-point topology of coronal jets in the open field of coronal holes. Sustained breakout reconnection is key to removing all of the overlying magnetic flux, allowing the erupting flux rope to reach the breakout current layer and reconnect with the external field. Furthermore, the quasi-uniform strength of the open field strongly suppresses the expansion of the flux rope during its rise. Generalisations of this model with varied inclinations of the open field \citep{Wyper2018,Wyper2019} and manner of energisation \citep{Wyper2018b} have revealed these features of the evolution to be quite general. In certain cases, coupling of the breakout feedback mechanism to an ideal instability of the flux rope was found to initiate the eruption \citep{Wyper2019}. 

The internal magnetic structure and evolution with height of the broader, fan-like pseudostreamer CMEs is less well understood. In \citet{Wyper2021}, we showed that if the pseudostreamer is topologically connected to a helmet streamer, then a coupled pseudostreamer/helmet-streamer blowout eruption can occur and produce a broad CME. However, that eruption was jet-like in the low corona, whereas most broad pseudostreamer CMEs appear to have a CME-like lifting-off of the flux rope at low heights. The lift-off is often accompanied by a rolling motion of the flux rope, e.g. \citet{Panasenco2011,Kumar2021}. Ultimately, the erupting flux rope is expected to reach the breakout current sheet where it will reconnect with the open field as in a jet; but when and where in the evolution this occurs is not well understood. Does the fan-like ejecta represent a flux rope still connected at both ends to the solar surface? Or has the flux rope reconnected after reaching a certain height? If so, where does the reconnection occur? 

In this paper we present a magnetohydrodynamic (MHD) simulation model of a fan-shaped pseudostreamer CME designed to address these questions. The setup is a generalisation of our model for coronal jets and reproduces many of the observed features of fan-shaped pseudostreamer CMEs. In \citet{Wyper2022}, we focused on the interchange reconnection dynamics of the early breakout process at the top of a model pseudostreamer. Here we have extended the run time of that simulation in order to investigate the subsequent eruption. 
Most importantly, we demonstrate that the magnetic evolution is exactly the same as that of mini-filament coronal jets. The model demonstrates that despite their differences in ejecta morphologies and scale, coronal jets and pseudostreamer CMEs belong to a continuum of eruptions unified by the pseudostreamer topology.

In \S 2 we describe the simulation setup. \S 3 gives an overview of the eruption and \S 4 describes the energy release and reconnection process in more detail. In \S 5 we discuss our results in the context of recent observations. Our conclusions are given in \S 6.

\begin{figure*}
    \includegraphics[width=1.0\textwidth]{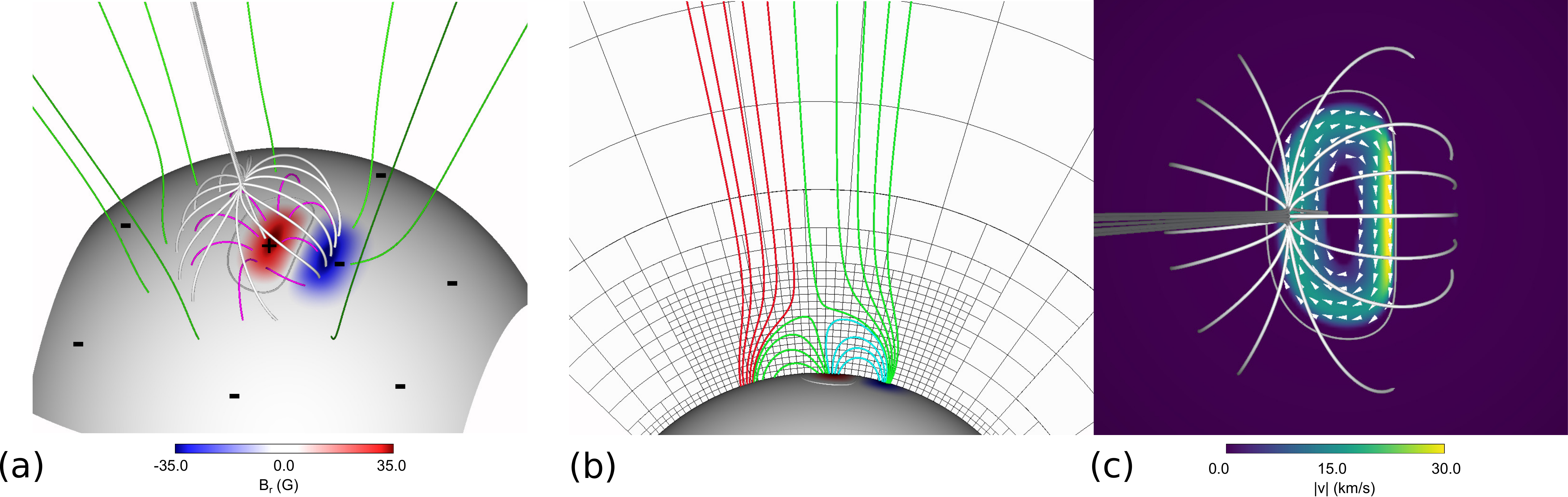}
    \caption{(a) The initial magnetic field. (b) Side view of the initial simulation grid blocks (each block contains $8\times8\times8$ grid cells).  (c) The surface driving profile. The PIL is shown in grey in each panel.}
    \label{fig:setup}
\end{figure*}

\section{Simulation setup}
The ideal, compressible MHD equations are solved by the Adaptively Refined Magnetohydrodynamics Solver \citep[ARMS;][]{Devore2008} in the following form:
\begin{align}
\frac{\partial \rho}{\partial t} + \grad \cdot \rho \vb &=0,  \\
\frac{\partial \rho \vb}{\partial t} + \grad \cdot \rho \vb \vb &= - \grad p +\frac{1}{4\pi}(\grad\times\Bb)\times\Bb + \rho \gb, \\
\frac{\partial \Bb}{\partial t} &= \grad\times(\vb\times\Bb), 
\end{align}
where $\rho$ is the mass density, $\vb$ the plasma velocity, and $\Bb$ is the magnetic field. Gravity takes the form $\gb = -GM_\odot/r^2\evr$. We assume that the plasma is an ideal gas with $p = 2(\rho/m_p)k_BT$, where $k_B$ is Boltzmann's constant and $m_p$ is proton mass. The temperature is assumed to be constant and uniform throughout the volume with $T = 1$\,MK.

The domain is a spherical wedge with radius $r\in [1R_\odot,20R_\odot]$ and latitude/longitude $\theta,\phi \in [-50.4^\circ,50.4^\circ]$. The magnetic field is initialised as a monopolar radial magnetic field of strength $b_0$ at $r = R_\odot$ together with $16$ sub-surface radially aligned dipoles, such that 
\begin{align}
\mathbf{B} = b_0\frac{R_\odot^2}{r^2}\mathbf{e}_r +  \sum_i M_i \left(\frac{d}{|\mathbf{r}-\mathbf{r}_i|} \right)^3[3(\mathbf{m}_i\cdot\mathbf{e}_r)-1]\mathbf{e}_r,
\end{align}
where $b_0 = -2.5$\,G, $R_\odot = 7\times10^{10}$\,cm, $\mathbf{m}_i$ is the unit vector in the direction of $\mathbf{r}-\mathbf{r}_i$, and $d = 8\times 10^9$\,cm. The values used for $M_i$ and $\mathbf{r}_i$ are given in Table \ref{tab:dipoles}. The field is shown in Figure \ref{fig:setup}(a) and takes the form of a bipolar surface flux distribution supporting a large-scale coronal null-point topology (null height $\approx 0.25\,R_\odot$ above the surface). The spacing between the dipoles and relative strengths of the dipoles and monopolar field closely matches our cartesian jet simulation model with vertical background field \citep{Wyper2018}, but on a much larger scale and in spherical geometry. The key difference between the two setups is the radial expansion of the monopolar field, which we will demonstrate plays a key role in the eruption evolution. 

\begin{table}[h]
\centering
  \begin{tabular}{|c|c|c|c|c|}
\hline
$i$ & $M_i$ & $r_i$ & $\theta_i$ & $\phi_i$\\ \hline
$1,2$ & $12$ & $6.2\times10^{10}$ & $+1.148^\circ$ & $\pm 6.548^\circ$     \\ 
$3,4$ & $12$ & $6.2\times10^{10}$ & $+1.148^\circ$ & $\pm 3.274^\circ$      \\ 
$5$ & $12$ & $6.2\times10^{10}$ & $+1.148^\circ$ & $0.0^\circ$     \\ 
$6,7$ & $12$ & $6.2\times10^{10}$ & $-2.126^\circ$ & $\pm 6.548^\circ$     \\ 
$8$ & $12$ & $6.2\times10^{10}$ & $-2.126^\circ$ & $0.0^\circ$     \\ 
$9,10$ & $-10.4$ & $6.2\times10^{10}$ & $-8.674^\circ$ & $\pm 6.548^\circ$     \\ 
$11$ & $-10.4$ & $6.2\times10^{10}$ & $-8.674^\circ$ & $0.0^\circ$     \\ 
$12,13$ & $-10.4$ & $6.2\times10^{10}$ & $-11.948^\circ$ & $\pm 6.548^\circ$     \\ 
$14,15$ & $-10.4$ & $6.2\times10^{10}$ & $-11.948^\circ$ & $\pm 3.274^\circ$     \\
$16$ & $-10.4$ & $6.2\times10^{10}$ & $-11.948^\circ$ & $0.0^\circ$     \\
\hline
  \end{tabular}
  \caption{Dipole parameters: $\mathbf{r}_i = (r_i,\theta_i,\phi_i)$; $M_i$ (G); $r_i$ (cm); and $\theta_i$,$\phi_i$ (degrees).}
\label{tab:dipoles}
\end{table}

Figure \ref{fig:setup}(b) shows the computational grid. A volume of fixed maximum refinement is centered around the closed field region. Outside of this volume, the grid refines adaptively as needed with the refinement criterion depending upon the local electric current density \citep{Karpen2012}. The base grid level was set to $16\times8\times8$ grid blocks (each block contains $8\times8\times8$ grid cells), with up to $4$ additional levels of refinement in this simulation \citep[$2$ fewer than in][where the aim was to track the evolution of small-scale plasmoids]{Wyper2022}. The atmosphere is initialised with a 1D isothermal \citet{Parker1958} wind solution and relaxed to a quasisteady state over $4\times10^4$\,s (see Fig.\ 1(b) of \citet{Wyper2022} for the resulting wind profile). All times in the rest of this paper are quoted from that point on, i.e., $t=0$  corresponds to the end of the relaxation and the start of the surface driving. 

The driving profile is the same as that used in our Cartesian jet setup \citep{Wyper2018} and is given by
\begin{align}
\mathbf{v}_{\perp}&= v_0 g(B_r)\mathbf{e}_r \times \boldsymbol{\nabla} B_r, \\
g(B_r) &=
\begin{cases}
   k_b\frac{b_r-b_l}{B_r}\tanh \left(k_b\frac{B_r-b_l}{b_r-b_l}\right), & b_l\leq B_r \leq b_r, \\
   0, & \text{otherwise}
\end{cases}
\end{align}
where $B_r$ is the normal field component on the lower boundary, $b_r = 30$ and $b_l = 1.6$ define the contours of $B_r$ within which the flow is restricted, and the constants $k_b$ and $v_0$ are set to $5$ and $3.079\times10^{13}$, respectively. By design the flow follows the contours of $B_r$, so it does not change the $B_r$ surface distribution. The spatial profile is shown in Figure \ref{fig:setup}(c). The driving is ramped up over $1000$\,s, held constant for $24000$\,s, and then ramped down just before the onset of the CME (halting at $t=2.5\times10^{4}$\,s or $6$\,hr $57$\,min). The driving speed peaks at $v_\perp \approx 30$\,km s$^{-1}$ in the center of the surface bipolar flux distribution. This speed is chosen for numerical convenience. Although relatively fast compared to solar surface flows, the flow is still substantially sub-Alfv\'{e}nic and sub-sonic, hence the pre-eruption closed magnetic field evolves quasistatically.

\begin{figure*}
    \includegraphics[width=1.0\textwidth]{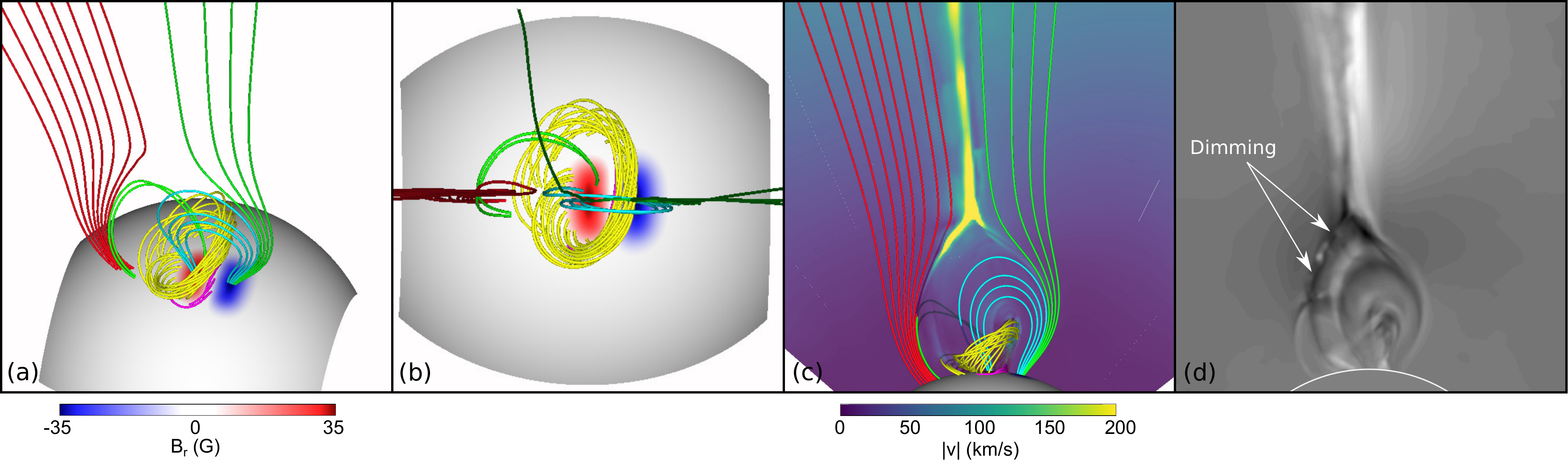}
    \caption{(a) and (b): Two views of the flux rope (yellow field lines) formed within the sheared filament channel at the end of the driving period ($t = 6$\,hr $57$\,min). Magenta field lines show short flare loop field lines. (c) $|v|$ showing the pre-eruption plasma jet (analyzed by \citet{Wyper2022}). (d) Synthetic white-light base-difference image ($7$\,hr $30$\,min - $6$\,hr $32$\,min) showing the pre-eruption dimming.}
    \label{fig:pre}
\end{figure*}

\section{Overview of the eruption}
\label{sec:overview}

\begin{figure*}
    \includegraphics[width=1.0\textwidth]{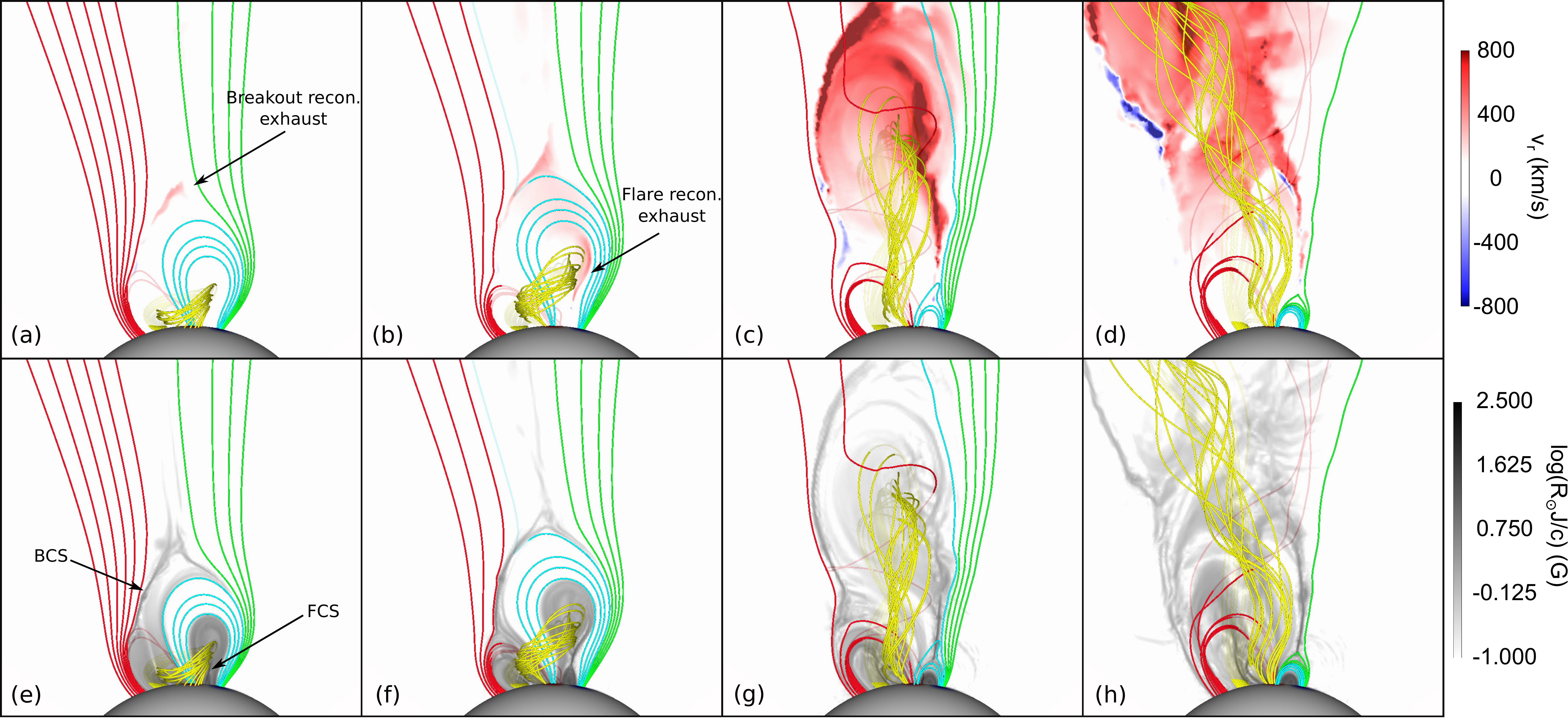}
    \caption{Top panels: radial velocity, $v_r$. Bottom panels: the logarithm of normalised current density $R_\odot |J|/c$. (a,e): $t=7$\,hr $38$\,min. (b,f): $t=8$\,hr $37$\,min. (c,g): $t=9$\,hr $10$\,min. (d,h): $t=9$\,hr $35$\,min. BCS = breakout current sheet; FCS = flare current sheet. An animation of panels (e) to (h) is available {showing the formation and eruption of the flux rope and the subsequent interchange reconnection. Key features are highlighted in the static figure}. The duration is 8\,s and runs from $t=0$ to $16$\,hr $23$\,min.}
    \label{fig:panels}
\end{figure*}

\subsection{Pre-eruption Reconnection}
The evolution during the driving phase is similar to that in the jet model, wherein the surface driving forms a quasi-circular filament channel overlying the PIL, Figure \ref{fig:pre}(a)-(b). The closed field expands asymmetrically and stresses the null point, forming a breakout current layer and inducing interchange reconnection there. The reconnection is resolved well enough that plasmoids form in the breakout current layer, and the reconnection eventually enters a bursty regime. This launches a plasma jet modulated by plasmoid ejections (Fig.\ \ref{fig:pre}(c)) and torsional Alfv\'{e}nic waves into the solar wind \citep{Wyper2022}. Aside from the jet itself, another observable signature of the onset of breakout reconnection is a dimming in synthetic white-light  base-difference images shown in Figure \ref{fig:pre}(d). {This is consistent with our findings of EUV dimmings in previous observational \citep{Kumar2021} and modeling \citep{Wyper2021} work.}

The breakout reconnection at this point is self-sustaining due to feedback between the outward expansion of the filament channel and the removal of overlying strapping field by the breakout reconnection. Well before this time, a twisted flux rope formed within the filament channel. This also occurred in the jet simulations: it arises from gradients in the surface driving profile creating a thin current layer inside the channel. Tether-cutting reconnection in this layer forms the twisted flux rope as part of a hyperbolic flux tube (HFT), but it does not trigger the eruption (see \S\ref{sec:recon} for further discussion).

\subsection{Eruption Onset and Early Evolution}
Figure \ref{fig:panels} shows the onset and early development of the eruption, which is triggered at $t \approx 8$\,hr when the flux rope begins to rise rapidly. This coincides with a transition from slow to rapid reconnection at the HFT and the formation of a vertical exhaust outflow, Figure \ref{fig:panels}(b,f). We denote this time (8 hr 37 min) as the onset of fast flare reconnection. Unlike the jet model, in which rapid reconnection is triggered only when the flux rope reaches the breakout current layer, in this case substantial strapping field remains above the flux rope when the fast reconnection turns on. The strapping field is carried out along with the flux rope as it continues to rise and accelerate, in the manner of a typical breakout CME \citep[e.g.,][]{Antiochos1999,Lynch2008}. As is typical in pseudostreamer CMEs \citep[e.g.,][]{Panasenco2011,Lynch2013,Sahade2022}, the erupting flux rope deflects toward the null-point breakout current sheet where the magnetic field strength is lowest, Figure \ref{fig:panels}(c,g). This deflection plus the exhaust jet from the reconnecting flare current sheet combine to create a rolling motion of the rising flux rope, Figure \ref{fig:panels}(c,d). See also the animation of this figure.

\begin{figure}
    \includegraphics[width=0.5\textwidth]{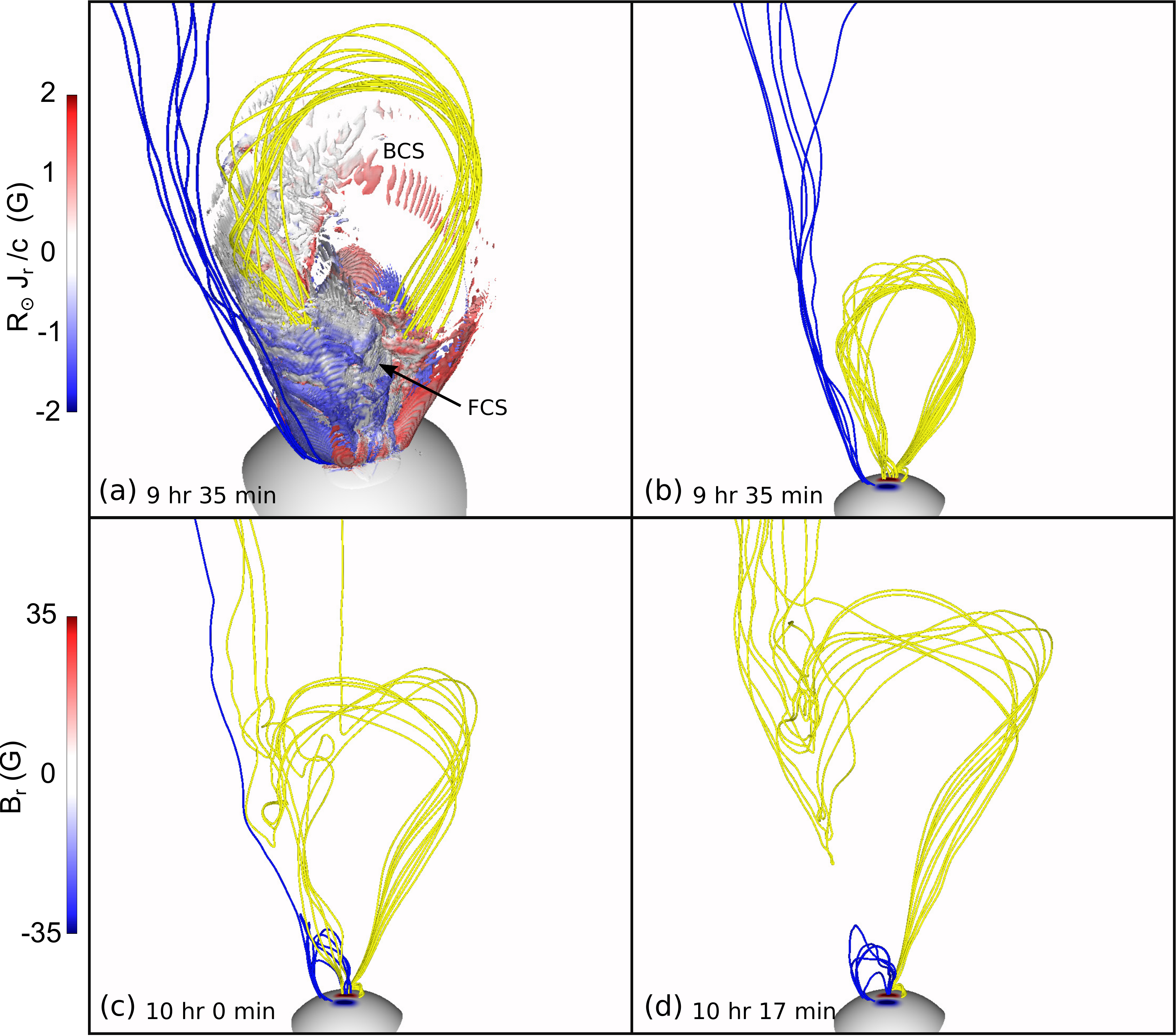}
    \caption{(a) Isosurface of normalised current density magnitude ($R_\odot J/c = 1.5$\,G) shaded by the radial component ($R_\odot J_r/c$). (b)-(d) Field line evolution showing the flux rope disconnection. An animation of panels (b) to (d) is available {showing the dynamic evolution of the field lines}. The duration is 1\,s and runs from $t=7$\,hr $55$\,min to $10$\,hr $33$\,min.}
    \label{fig:disc}
\end{figure}

\begin{figure}
    \includegraphics[width=0.5\textwidth]{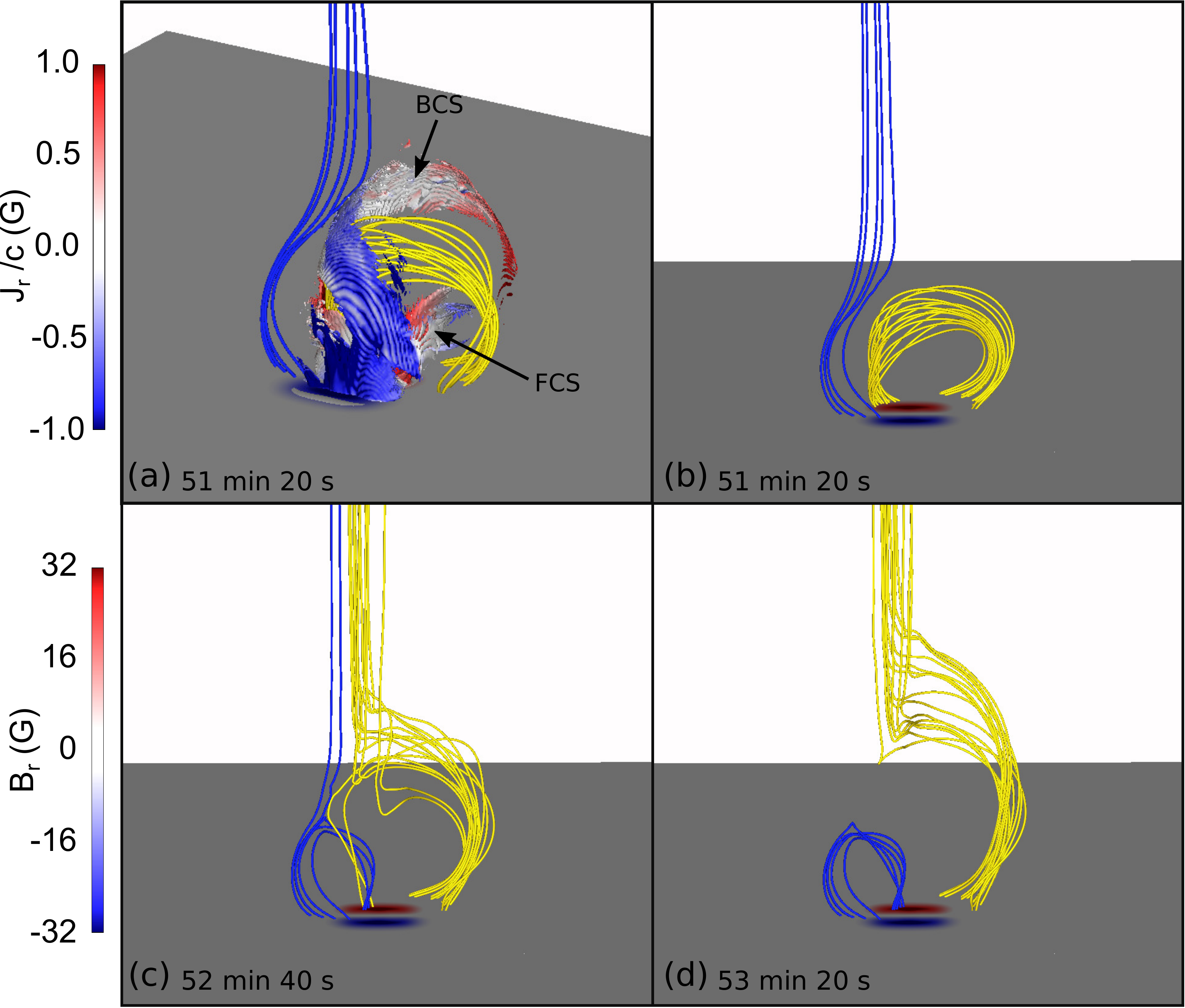}
    \caption{Flux rope disconnection in the jet simulation. (a) Isosurface of normalised current density ($J/c = 1.2$\,G) shaded by the radial component ($J_r/c$). (b)-(d) Field line evolution showing the flux rope disconnection.}
    \label{fig:jet}
\end{figure}

\begin{figure}
    \includegraphics[width=0.5\textwidth]{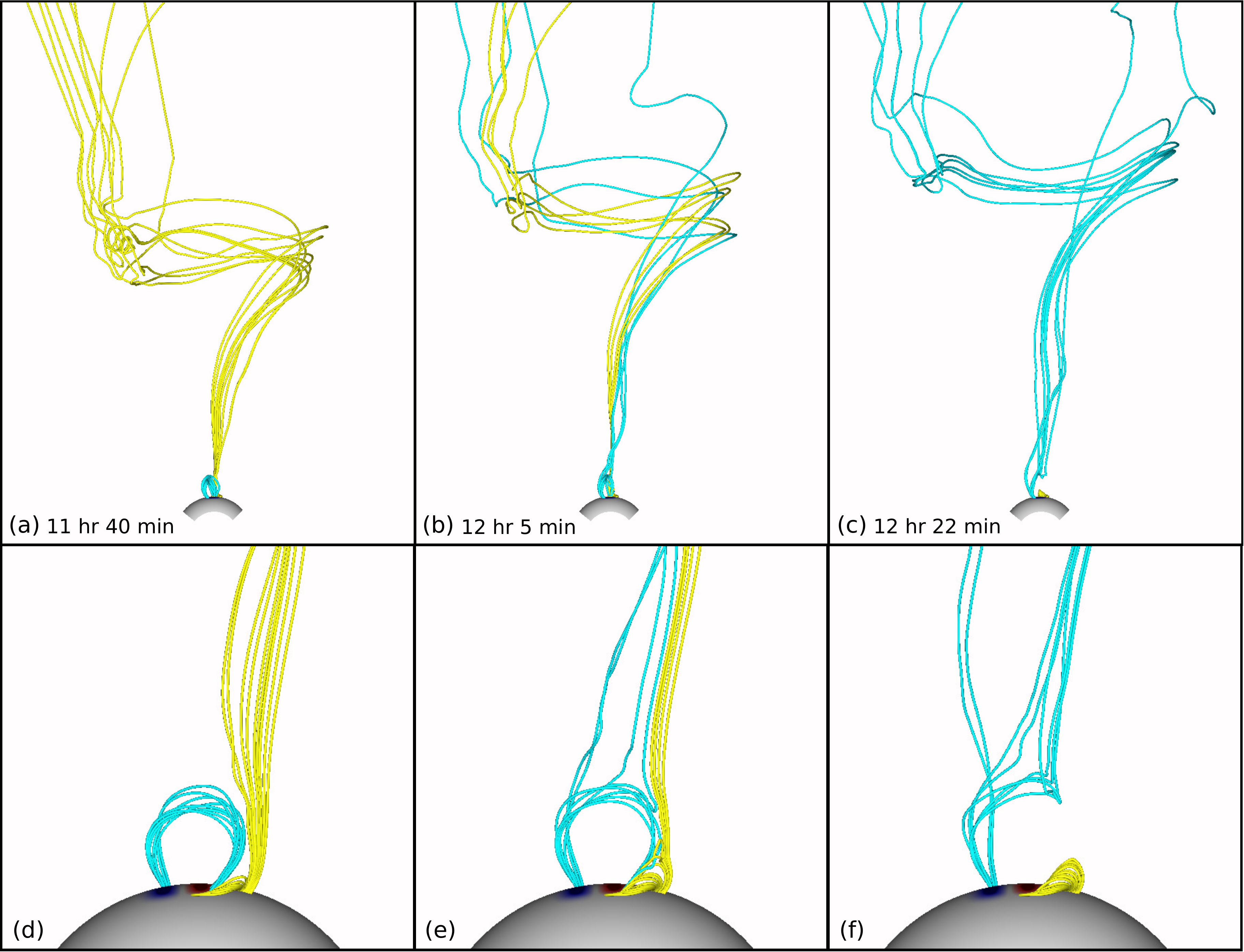}
    \caption{Second stage of CME flux rope disconnection.  {The top panels show field lines extending into the CME traced from the surface. The bottom panels show a close-up view (from the side) of the same field lines at the same times showing the re-closing of the yellow field lines and the shift in the CME flux rope footpoints to the cyan field lines.}}
    \label{fig:discon2}
\end{figure}

\begin{figure*}
    \includegraphics[width=1.0\textwidth]{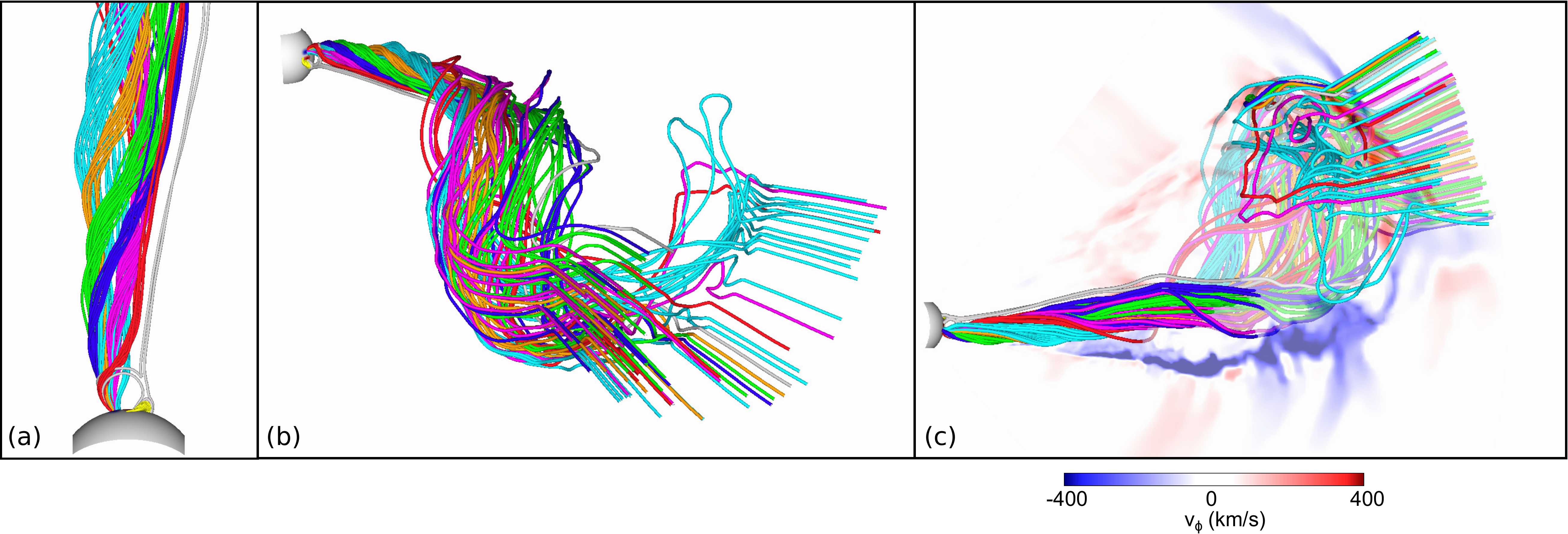}
    \caption{Flux ropes formed following the second phase of interchange reconnection at $t = 12$\,hr $30$\,min. Light blue/cyan shows the flux tube with the twist of the original rope. Other twisted flux tubes are formed from plasmoids ejected as part of the bursty interchange reconnection process. (a) Close-in view of the flux ropes; note that the yellow field lines showing the original rope footpoints are now closed. (b) Farther-out view showing how the flux ropes wrap into the CME structure. (c) Cut showing the large-scale rotation of the field lines.}
    \label{fig:ropes}
\end{figure*}

\begin{figure*}
    \includegraphics[width=1.0\textwidth]{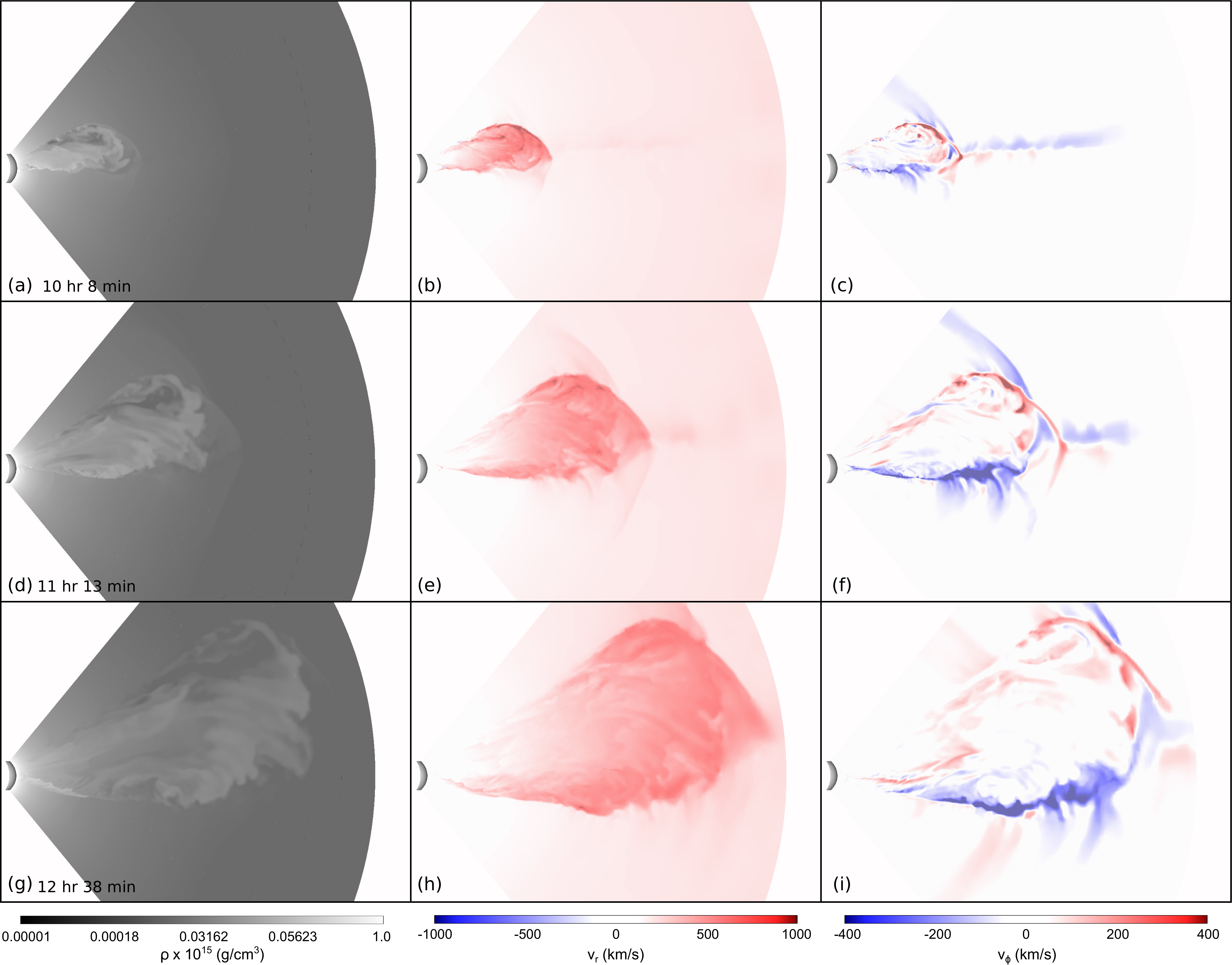}
    \caption{$\rho$ (left), $v_r$ (middle), and $v_\phi$ (right) in a cut ($\phi=6^\circ$) through the CME at different times.}
    \label{fig:cme}
\end{figure*}

\subsection{Flux Rope Disconnection}
As the flux rope rises, the breakout reconnection continues to erode the strapping field while the flare reconnection continues to strengthen the flux rope. The null point and breakout current layer are pushed out to several solar radii before the strapping field is exhausted and the flux rope itself starts to reconnect. Despite the greatly expanded size of the open-closed separatrix surface, the flux rope reconnection proceeds in exactly the same manner as in the jet simulation. The breakout and flare current layers combine into one long current sheet that wraps around the flux rope, Figure \ref{fig:disc}(a). The null point then moves within this sheet, sliding from the top of the separatrix and down the side to ultimately end up below the flux rope. At the end point, the flux rope has reconnected completely and is comprised entirely of open field lines. 

Figure \ref{fig:disc}(b) to (d) shows the field line evolution during this phase. The majority of the flux rope disconnection occurs when the null point (i.e., the main interchange reconnection site) is on the side/flank of the flux rope rather than at the apex as one might anticipate. As the flux rope expands outwards and the legs become radially oriented, one of the legs will have field lines anti-parallel to the adjacent coronal hole field lines, shown in blue in Figure \ref{fig:disc}. As a result, there must be a current layer separating these field regions and, in general, flux rope disconnection will occur along that leg of the CME. As shown in Figure \ref{fig:disc}(d), this leads to the formation of a transient V-shape in the flux rope shortly after disconnection. This kink straightens out as the flux rope continues to rise (see the animation of this figure). 

For comparison, Figure \ref{fig:jet} shows the combined current sheets and flux rope disconnection in the Cartesian jet simulation with vertical background field \citep{Wyper2018}. As this flux rope is less expanded, the disconnection occurs nearer its apex, but otherwise the disconnection process proceeds in exactly the same manner. 

\subsection{CME Magnetic Structure}
In the wake of the flux rope disconnection, interchange reconnection continues to sequentially open sheared closed field lines while closing down unsheared open field lines. This process progresses around the circular PIL (see \S\ref{sec:con}). Ultimately, the closed-field footprint returns to approximately where it started. As part of this subsequent evolution, the erupting flux rope undergoes a second leg reconnection event. However, this one is much less dramatic, occurring much closer to the surface and taking the form of a shift in the flux rope footpoints, Figure \ref{fig:discon2}. The yellow field lines of the erupted flux rope close down, while the neighboring closed cyan field lines open up. In effect, the erupting flux rope footpoints shift from one side of the pseudostreamer to the other (note the change in color of the propagating CME field lines from yellow to cyan). This connectivity of the field lines in the CME is maintained thereafter.

The magnetic structure of the CME as a whole, on the other hand, is more complex. As in the breakout reconnection phase, the interchange reconnection during the disconnection phase is also bursty and dominated by plasmoid ejection. This dynamically creates further twisted flux tubes within the null-point current layer that propagate into the open field in the wake of the disconnection. Figure \ref{fig:ropes} shows a selection of these flux tubes. The original erupting flux rope is shown in cyan, while plasmoid flux tubes are shown in the order in which they were launched sequentially: orange, green, blue, pink, and then red. The twist within the flux tubes propagates as torsional Alfv\'{e}n waves behind the main CME, while their field lines map into the main body of the CME and thread through the sheath that surrounds the original flux rope, Figure \ref{fig:ropes}(b) and (c). 

Following the two-stage disconnection and reconnection of the original coronal flux rope, the erupting structure reaches its final magnetic configuration. The main body of the CME contains an embedded twisted flux tube that is the remnant of the original erupting flux rope. This is surrounded by a sheath of highly distorted but untwisted field. Following behind in the trailing wake of the CME are bursty outflows and twisted flux tubes launched by the interchange reconnection process. All of these structures are comprised of open field lines.

\subsection{CME Plasma Evolution}
The evolution of some additional properties of the CME are shown in Figure \ref{fig:cme}. The embedded flux rope (top of the CME in this view) is visible as a region of depleted density, Figure \ref{fig:cme}(a). Outside of this feature, the CME appears broadly unstructured in nature. In a follow-up paper we will make a detailed exploration of the white-light properties of the CME structure from different viewing angles.

Figures \ref{fig:cme}(c), (f), and (i) show that the main body of the CME rotates as it propagates. It is tempting to equate this rotation to that of {the outflowing plasma about the spire in} helical coronal jets resulting from minifilament/filament channel eruptions. However, this rotation is {qualitatively} different. In the case of jets it is the untwisting of the erupted flux rope {reconnected onto open field lines} that drives the rotation. Here, the rotation is driven by the whip-like motion of the flux rope {axis} and the sheath of field lines surrounding it, Figure \ref{fig:ropes}(b). It is the precessing axis of the flux rope {-- i.e., the evolution of the writhe --} that drives the large-scale rotation, rather than its twist. {Due to the expanded horseshoe-like shape of the flux rope prior to disconnection, the axis of the twisted open flux tube created by the disconnection is highly kinked. The straightening out of the axis, along with the straightening out of the surrounding sheath field lines, creates the large-scale rotation of the CME.} This evolution is actually closer in nature to the whip-like field-line motion in the helical jet model of \citet{Pariat2009} for jets without filament channels. The twist of the flux rope also propagates along its axis, as is true of the plasmoid-generated flux tubes in the wake of the CME. In this sense there are torsional waves within waves involved in the CME evolution.

Also notable is the MHD shock launched ahead of the CME by the initial flux rope expansion and the reconnection outflow from the breakout reconnection. The latter is most visible in the $v_\phi$ component, Figure \ref{fig:cme}(c), and is eventually overtaken by the CME at later times. \citet{Wyper2021} and \citet{Kumar2021} noted similar pre-eruption jets in their simulation and observational studies, respectively, of pseudostreamer CMEs. 

\begin{figure}
    \includegraphics[width=0.5\textwidth]{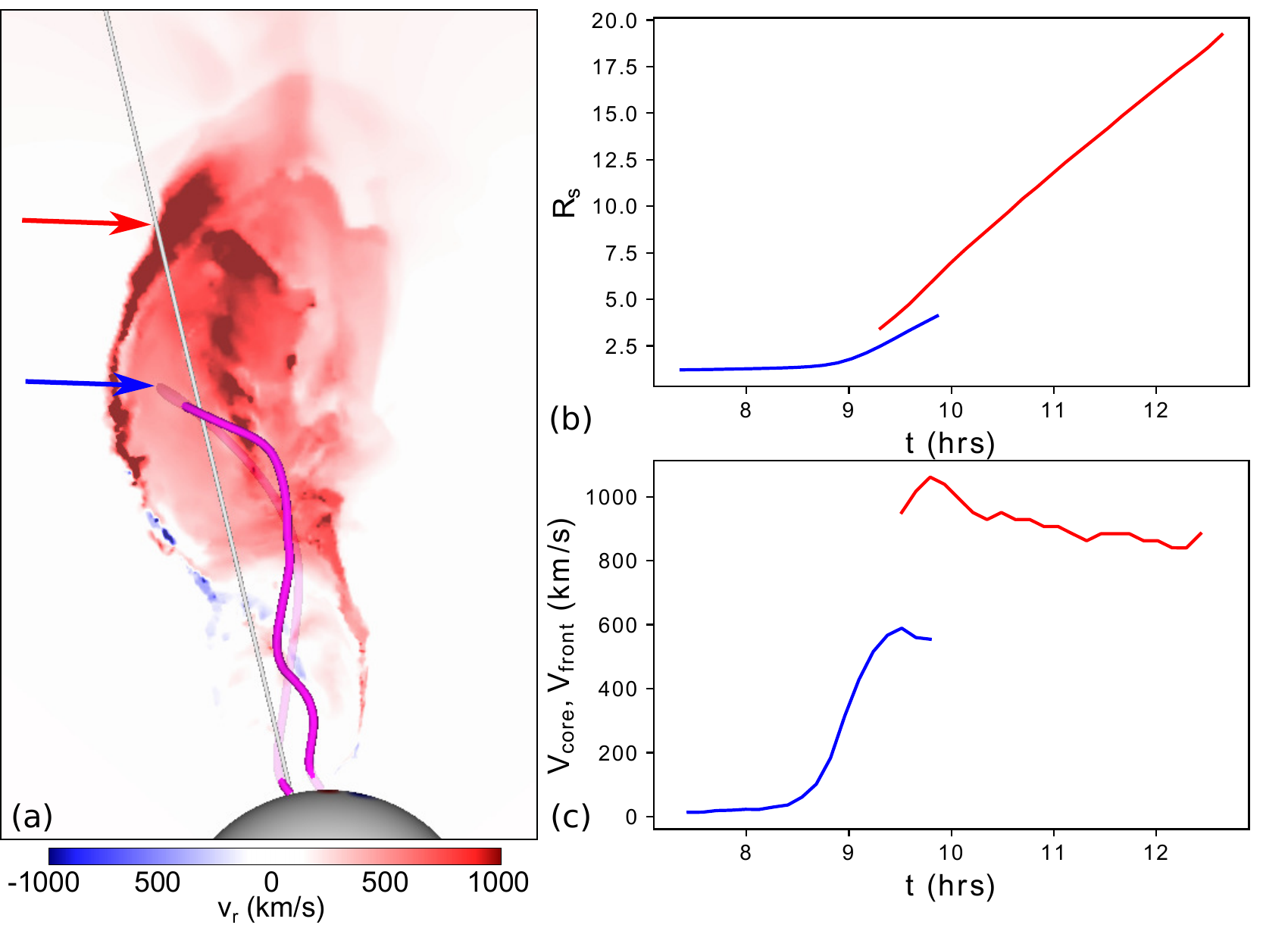}
    \caption{(a) Shading shows $v_r$ in the plane $\phi=0$ at $t = 9$\,hr $35$\,min. The white line shows the path along which the CME front speed is calculated. The red arrow shows the position of the front at this time. The purple field line shows the flux rope axis. The blue arrow shows the highest point of this field line. (b) Radial positions of the front (red) and highest point of the axis field line (blue). (c) Radial speeds: $V_{front}$ (red), $V_{axis}$ (blue).}
    \label{fig:vcme}
\end{figure}

In Figure \ref{fig:vcme} we show two measures of the speed of the ejecta. The initial rise of the flux rope is well captured by following the highest point on a field line (purple) traced through the axis of the rope. Initially the flux rope accelerates before plateauing at $v \approx 600$\,km s$^{-1}$ once the disconnection begins; panel (c), blue curve. Beyond this point the changing connectivity of the CME makes it difficult to follow individual magnetic field structures. To estimate the overall CME speed beyond this time, a radial sample was taken (white line; panel (a)) and the shock front was identified. Height/time and speed curves for the front are shown in red in panels (b) and (c). Evidently, once the disconnection begins the acceleration of the erupting material ceases, with the CME front propagating at a nearly constant speed thereafter. Qualitatively, the height/time plot in panel (b) closely matches the height/time plots of essentially all pseudostreamer CMEs \citep[e.g.,][]{Wang2015}. {The high speed of our explosive, fast CME is near the upper end of the 250-700 km s$^{-1}$ range observed.}

\section{Energy release vs.\ reconnection}
\label{sec:recon}

\begin{figure*}
    \includegraphics[width=1.0\textwidth]{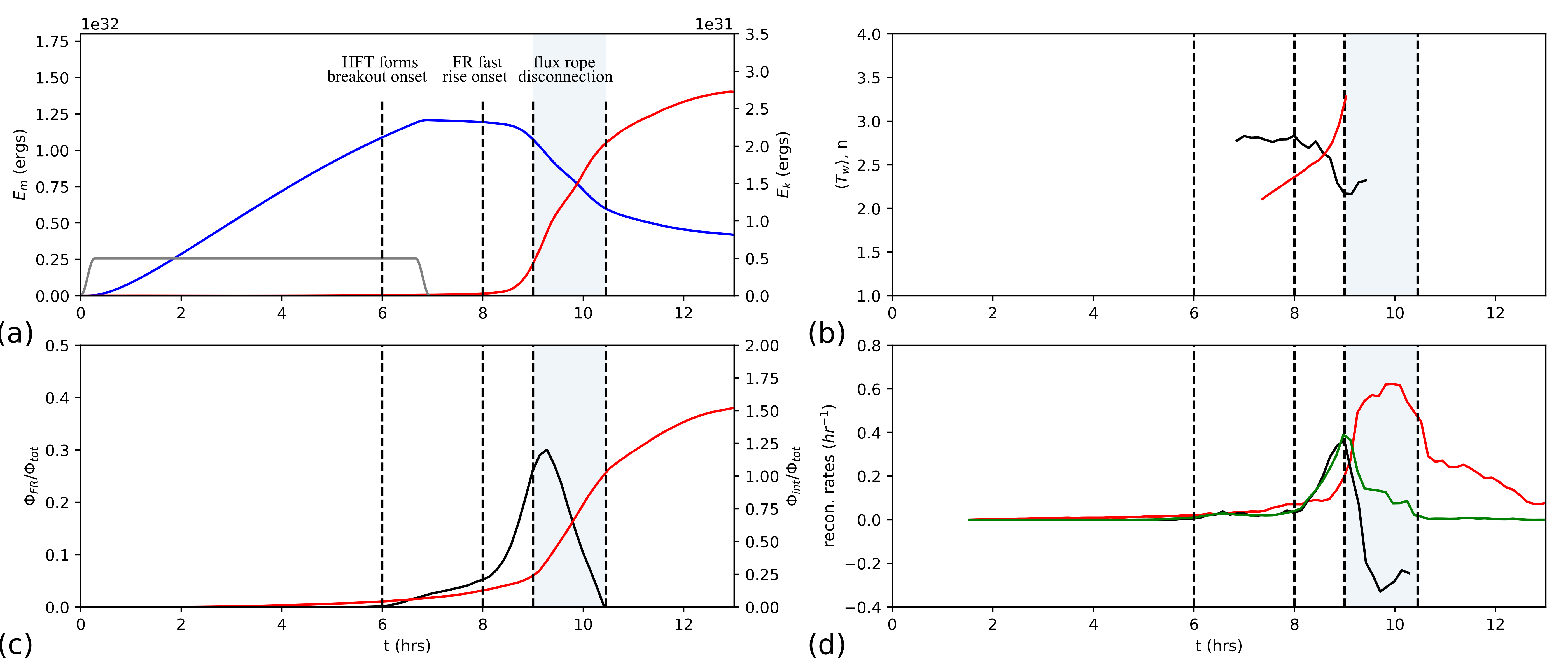}
    \caption{(a) Free magnetic ($E_m$, blue) and kinetic ($E_k$, red) energies. Grey: Time profile of the driving (scaled to fit on the plot). Blue shading highlights the flux rope disconnection period. (b) Average number of turns within the flux rope ($\langle T_w \rangle$, black) and the decay index at the flux rope axis ($n$, red). (c) flux rope flux ($\Phi_{FR}$, black) and cumulative interchange reconnected flux ($\Phi_{int}$, red), normalised by the total closed-field flux ($\Phi_{tot}$). (d) Rates of change of these normalised fluxes (same colour scheme). Green: Normalised closed-field reconnection rate obtained by calculating the swept-over flare-ribbon flux (see text for details). HFT = hyperbolic flux tube; FR = flux rope. 
    }
    \label{fig:energies}
\end{figure*}

\begin{figure}
    \includegraphics[width=0.5\textwidth]{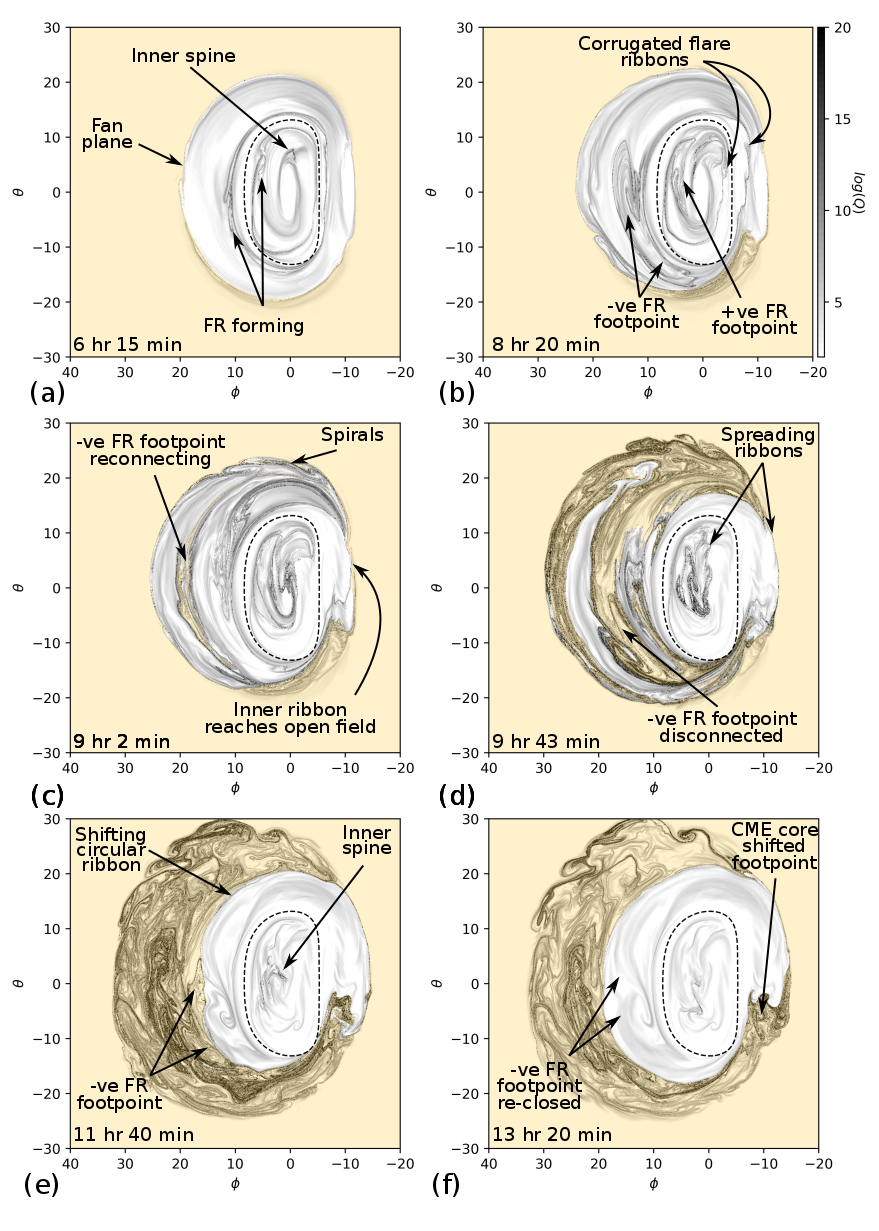}
    \caption{Evolution of the squashing factor ($Q$) on the surface throughout the evolution. Yellow shading shows the open field regions. An animation is available showing the {motion of the QSLs and the closed-field region across the surface. The static figure highlights the main features in 6 panels.} The duration is 4\,s and runs from $t=4$\,hr $52$\,min to $13$\,hr $3$\,min.}
    \label{fig:Q}
\end{figure}

\begin{figure}
    \includegraphics[width=0.5\textwidth]{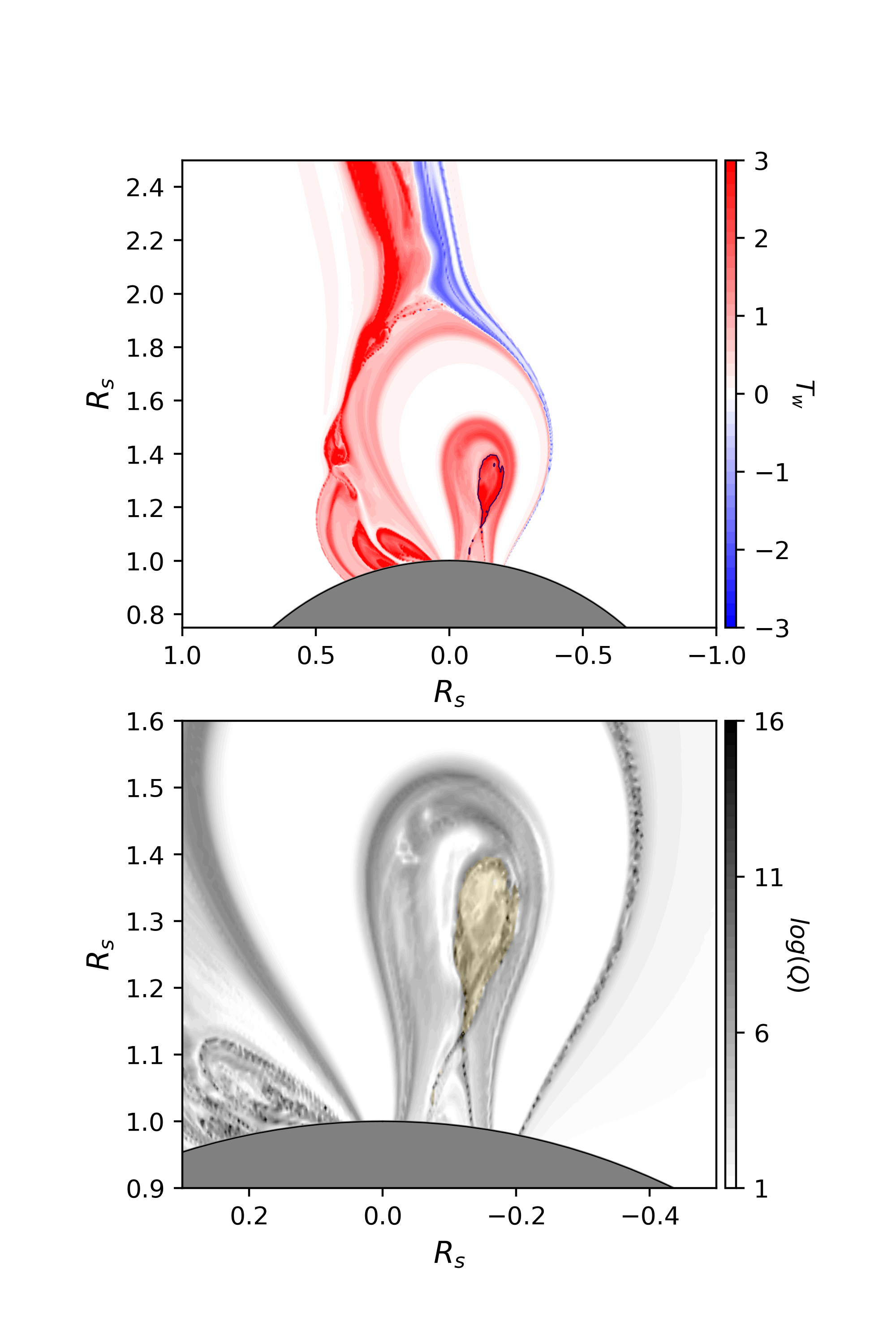}
    \caption{Top: Turns parameter $T_w$ at $t=8$\,hr $12$\,min; the black contour shows the length-weighted twist parameter $\tau_w=6$ identifying the flux rope. Bottom: $Q$; the flux rope is shaded yellow.}
    \label{fig:tw}
\end{figure}

\subsection{Energetics}
Having summarised the main evolutionary features of the eruption, we now explore the manner of the energy release and its relation to reconnection. To do so, we first define the free magnetic and kinetic energies of the system to be
\begin{align}
E_m &= \iiint_{V} \frac{1}{8\pi} \mathbf{B}^2 \,dV - \left(\iiint_{V} \frac{1}{8\pi} \mathbf{B}^2 \,dV \right)_{t=0}, \\
E_k &= \iiint_{V} \frac{1}{2}\rho \mathbf{v}^2 \,dV - \left(\iiint_{V} \frac{1}{2}\rho \mathbf{v}^2 \,dV\right)_{t=0}.
\end{align}
Time $t=0$ corresponds to the end of the relaxation period when both quantities have stopped varying to within a few percent. Their subsequent evolution is shown in Figure \ref{fig:energies}(a). The early stages of the simulation ($t<6.5$\,hr) are marked by a gradual increase in free magnetic energy as the closed field is sheared. By contrast there is negligible additional kinetic energy in the system, including when the HFT forms. It is clear that the HFT formation in our simulation is a low-energy process, not an explosive one. This differs from most previous breakout CME studies where the eruption commences when flare reconnection begins and the HFT first forms \citep[e.g.,][]{Karpen2012}.

The start of the breakout phase overlaps with the end of the driving phase. Breakout reconnection starts at $t\approx 5$\,hr, but becomes more rapid and self-sustaining around the time the HFT forms at $t\approx 6$\,hr. Once the driving ceases, this leads to a steady, slow decrease in $E_m$ and a small increase in $E_k$. The evolution switches to a rapid increase in $E_k$ and drop in $E_m$, characteristic of a breakout CME-like evolution, in the early phase of the eruption ($t\approx 8$\,hr to $t \approx 9$\,hr). The near-exponential rise in $E_k$ (and drop in $E_m$) then slows slightly throughout the flux rope disconnection, before tapering off once the disconnection finishes. By comparison, the disconnection in the jet simulation occurs much more rapidly and initiates the rapid rise in $E_k$ (and drop in $E_m$). Following the disconnection of our pseudostreamer CME flux rope, $E_k$ continues to rise as more mass is ejected and the drop in $E_m$ tapers off as the interchange reconnection relaxes the closed field toward a new equilibrium. In this end state, the closed field still retains a small amount of free magnetic energy, as the opening/closing process is not $100$\% efficient in transferring the free energy and helicity to the open field. This is another general property of jet-like eruptions \citep[e.g.,][]{Pariat2009,Wyper2016,Karpen2017,Wyper2018}. 

\subsection{Surface Connectivity Evolution}
\label{sec:con}
To relate the energy release to the reconnection process, it is instructive to consider how the surface connectivity changes during the eruption. Figure \ref{fig:Q} shows the evolution of the squashing factor $Q$ \citep{Titov2007} (grey scale) on the surface throughout the simulation. The squashing factor shows the surface imprint of magnetic (quasi-)topological boundaries in the volume, and has been shown to closely correlate with flare ribbons in observed flares \citep[e.g.,][]{Janvier2013,Savcheva2016}. To calculate $Q$ we implemented the method of \citet{Tassev2017} on adaptive grids and applied it to the data from ARMS.

Figure \ref{fig:Q}(a) shows the surface connectivity near the end of the driving period. The fan plane footprint is a closed ring of $Q$ at the boundary between open (yellow shading) and closed field regions. The footpoint of the inner spine is also highlighted. Two small hooks of $Q$ denote the formation of the first flux rope field lines (and simultaneously the HFT) in the simulation volume. Figure \ref{fig:Q}(b) shows how this has evolved by the the early stages of the eruption. The flux rope footpoints have grown considerably and the hooked ends of the quasi-separatrix layer (QSL) ribbons are now clearly discernible. Furthermore, the hooks have spread farther around the polarity inversion line (PIL) and two parallel ribbons are spreading out from the right side of the PIL (see also the animation). Both patterns follow from reconnection at the HFT below the flux rope. In fact, this ribbon evolution is exactly that of the standard two-ribbon flare model \citep[e.g.,][]{Aulanier2012,Janvier2013}, here embedded within the spine-fan topology of the pseudostreamer for a filament channel formed above a circular PIL. Additionally, the straight sections of the ribbon exhibit a corrugated structure, which is the imprint of plasmoid flux ropes formed within the flare current layer \citep{Wyper2022,Dahlin2022}. 

Figure \ref{fig:Q}(c) shows the beginning of the flux rope disconnection, whichs occurs when the breakout and flare current sheets combine into one long sheet, Figure \ref{fig:disc}(a). The corresponding imprint of this merger in the QSL ribbons is when the two inner ribbons reach the circular ribbon (on the negative side) and the inner spine (on the positive side). The two ribbon systems meet as there is now no intervening flux between the erupting flux rope and the open/closed boundary. The ``flare'' reconnection at this point becomes interchange reconnection, and the null point moves below the flux rope. This is exactly the same surface-connectivity evolution seen in our jet model \citep{Pariat2023}. Furthermore, as in the jet simulation the formation of plasmoids in the null-point current layer imparts spiral structure to the circular ribbon \citep[e.g.,][]{Pontin2015,Wyper2016b}. 

The negative footpoint of the erupting flux rope now rapidly opens up while the ribbons on the other side of the closed-field region continue to spread apart, Figure \ref{fig:Q}(d). In the aftermath of the disconnection, the interchange reconnection continues around the PIL, with the fan plane first shifting up (Fig.\ \ref{fig:Q}(e)), and then left and down (Fig.\ \ref{fig:Q}(f)); see also the animation of this figure. This latter shift closes the field back down over the negative flux rope footpoint, moving the CME footprint as shown in Figure \ref{fig:discon2}. Ultimately, this puts the closed-field region roughly back where it was before the eruption, with the surrounding open flux now having received most of the twist/helicity that was injected into the closed field by the driving. Therefore, the footpoints of the open field lines threading the CME are adjacent to the closed field, and are seen as a broad region of complex $Q$ structure in the open field, Figure \ref{fig:Q}(f). Analogous features form in the jet simulations \citep{Wyper2016b}. 

\subsection{Flux Rope Identification}
To complement the evolution of reconnection in the system inferred from the surface connectivity, we also isolated the flux rope itself. After some experimentation, we found that a reliable non-dimensional field-line-integrated quantity for identifying the flux rope is 
\begin{align}
\tau_w = T_w \frac{L}{L_{\rm PIL}},
\end{align}
where $T_w$ is defined as \citep[][]{Berger2006,Liu2016} 
\begin{align}
T_w = \int_{L} \frac{\boldsymbol{\nabla}\times\mathbf{B}\cdot\mathbf{B}}{4\pi |\mathbf{B}|^2} \,dl,
\end{align}
and $L$ and $L_{\rm PIL}$ are, respectively, the field-line length and the length of the PIL along which the flux rope forms (here this is the entire circular PIL). $T_w$ is the average number of turns of neighbouring field lines around the given field line; it reduces to evaluating the force-free parameter on the field line if the field is locally force-free. Once the eruption gets underway, the field is far from force free and includes many small-scale flux ropes within the breakout current sheet. The weighting of $L/L_{\rm PIL}$ ensures that the main coronal flux rope is preferentially identified. Furthermore, we limit the calculation of turns $T_w$ to closed field lines with well-defined, finite lengths $L$. This procedure enables us to identify flux-rope field lines unambiguously by monitoring $\tau_w$ prior to their disconnection. 

Figure \ref{fig:tw}(a) shows the flux rope identified using this procedure in a vertical cut; the threshold $\tau_w=6$ is contoured in black. Figure \ref{fig:tw}(b) shows that this threshold fully captures the flux rope (shown in yellow), which in this plane should be contained within a closed loop of high $Q$ above the HFT \citep[e.g.,][]{Savcheva2016}. {Noting from Figure \ref{fig:tw} that the average twist within the contoured region is $T_w \approx 3$, this indicates that the flux rope field lines are roughly twice the length of the PIL, $L/L_{\rm PIL} \approx 2$.} 

\subsection{Reconnection Rates}

Figure \ref{fig:energies}(c) shows the increase in the toroidal magnetic flux ($\Phi_{FR}$) contained within the flux rope versus time, calculated by integrating the field component through the plane over the region identified in Figure \ref{fig:tw}. The value is normalised by the total flux ($\Phi_{tot}$; all closed) of the minority polarity. The plot shows that the flux rope contains as much as about $30$\% of the magnetic flux within the closed field. By comparison, when we apply the same analysis to the vertical jet from \citet{Wyper2018}, the flux rope accumulates a maximum of only about $10$\% of the closed flux. This explains, at least in part, why the erupting CME flux rope takes a comparatively long time to disconnect: three times the amount of flux must be processed. 

The magnetic flux cumulatively reconnected by interchange reconnection ($\Phi_{int}$) is shown in red in Figure \ref{fig:energies}(c) for comparison. This shows the total flux that is opened (or closed) over time, again normalised by the total closed flux. By the time the flux rope leg has completely disconnected ($t$ just over $10$ hours), the cumulatively opened/closed flux equals the entire amount of closed-field flux; i.e., by this time all of the closed flux has likely interchange-reconnected once. The subsequent re-closing down of the opened field then grows the total $\Phi_{int}$ well past the amount of flux in the closed field.

The rates of change of the two quantities are shown in Figure \ref{fig:energies}(d). The red curve is the interchange reconnection rate, representing the breakout reconnection rate before the flux rope disconnects and the fast flare-like reconnection rate after. The black curve represents the rate of closed/closed flare reconnection occurring at the HFT prior to the flux rope disconnection. The negative rate after disconnection is an artifact of identifying only closed flux-rope field lines. The breakout reconnection starts out the fastest and is well underway when the fast flare reconnection is triggered at $t\approx 8$\,hr. Shortly after this time, the HFT reconnection rate rapidly surpasses the interchange reconnection rate. It is notable that the interchange reconnection rate does not increase significantly after onset of the rapid HFT reconnection and flux-rope rise around $t\approx 8$\,hr. This indicates that the fast HFT flare reconnection is driven by the rapid rise of the flux rope, rather than by an increase in the removal of strapping flux via breakout reconnection (which occurs in the jet simulation). The breakout reconnection rate increases later in reaction to the flux rope driving into the breakout current sheet from below, which initiates the disconnection of the flux rope from the surface.

The amount of flux contained within the flux rope also can be determined by calculating the surface flux swept out by the flare reconnection, in the manner of two-ribbon flares \citep[e.g.,][]{Kazachenko2022}. Field lines undergoing closed/closed flare reconnection were identified if their length changed more than $40\%$ from one time to the next. Care was taken to not double-count the flux and to exclude changes due to the interchange reconnection. The green curve in Figure \ref{fig:energies}(d) shows \emph{one-half} the normalised rate of flux swept out in the negative (majority) polarity closed-field footprint. The result closely matches the rate of toroidal flux accumulation within the flux rope (black curve) prior to disconnection. This close agreement suggests that there are equal increases in the poloidal and toroidal fluxes within the flux rope at this time. Qualitatively, it is consistent with the presence of a strong guide field within the erupting filament channel where the flux rope forms. Quantitatively, it shows that the fluxes swept out by the hooks and straight sections of the QSL ribbon are the same, as they are conjugate footpoints of the reconnecting field.

\subsection{Ideal Flux Rope Evolution}
Here we focus on the ideal evolution of the flux rope by averaging the twist on each field line within the region shown in Figure \ref{fig:tw}(b). This gives an approximation to the overall twist of the flux rope \citep[e.g.,][]{Liu2016}. The black curve in Figure \ref{fig:energies}(b) shows that the flux rope is highly twisted at formation, averaging $\langle T_w \rangle \approx 3$, or nearly three turns along its length. This is well into the unstable range of the kink instability \citep[e.g.,][]{Torok2005}. Notably, the number of turns begins to decrease after eruption onset at $t\approx 8$\,hr, suggesting that twist in the flux rope is being converted to writhe of its axis at this late stage. This result implies that the flux rope does, indeed, kink, but not before the eruption onset. 

The red curve in Figure \ref{fig:energies}(b) shows the decay index 
\begin{align}
n = -\frac{d(\ln(B_{\rm ex})}{d(\ln(h))},
\end{align}
where $h$ is the height above the lower boundary of the highest point on a field line approximating the flux rope axis and ${\bf B}_{\rm ex}$ is the field external to the flux rope. The decay index is difficult to calculate accurately; see \citet{Zuccarello2015} for an in-depth discussion. The index may be approximated by taking as the external field (${\bf B}_{\rm ex}$) the horizontal component of the potential field at the flux rope axis \citep{Zuccarello2015}, and we do so here. Figure \ref{fig:energies}(b) shows that the decay index starts at quite a high value $n \approx 2.0$ and increases steadily until reaching a value $n \approx 2.5$ when the eruption gets underway. This value is consistent with values in simulations of CMEs triggered by the torus instability \citep[e.g.,][]{Kliem2006,Aulanier2012}, although it is substantially higher than the value $n \approx 1.5$ often quoted as the threshold for instability.

The onset of fast flux rope acceleration  ($t \approx 8.5$\,hr) occurs well after the onset of self-sustaining breakout reconnection and flux-rope formation due to tether-cutting reconnection ($t \approx 6$\,hr). The flux rope survives an extended interval apparently unstable to the kink mode, as measured by its average twist $\langle T_w \rangle$, and to the torus mode, as measured by its decay index $n$. Taken together, our analysis suggests that neither mechanism is responsible for the transition from slow to fast rise and eruption onset in the pseudostreamer. In contrast, this transition is clearly concurrent with the onset of fast flare reconnection below the coronal flux rope, as evidenced by the abrupt turning up of the FR flux curve in Figure 10(c) and the FR reconnection rate curve in Figure 10(d).

\begin{figure*}
    \includegraphics[width=1.0\textwidth]{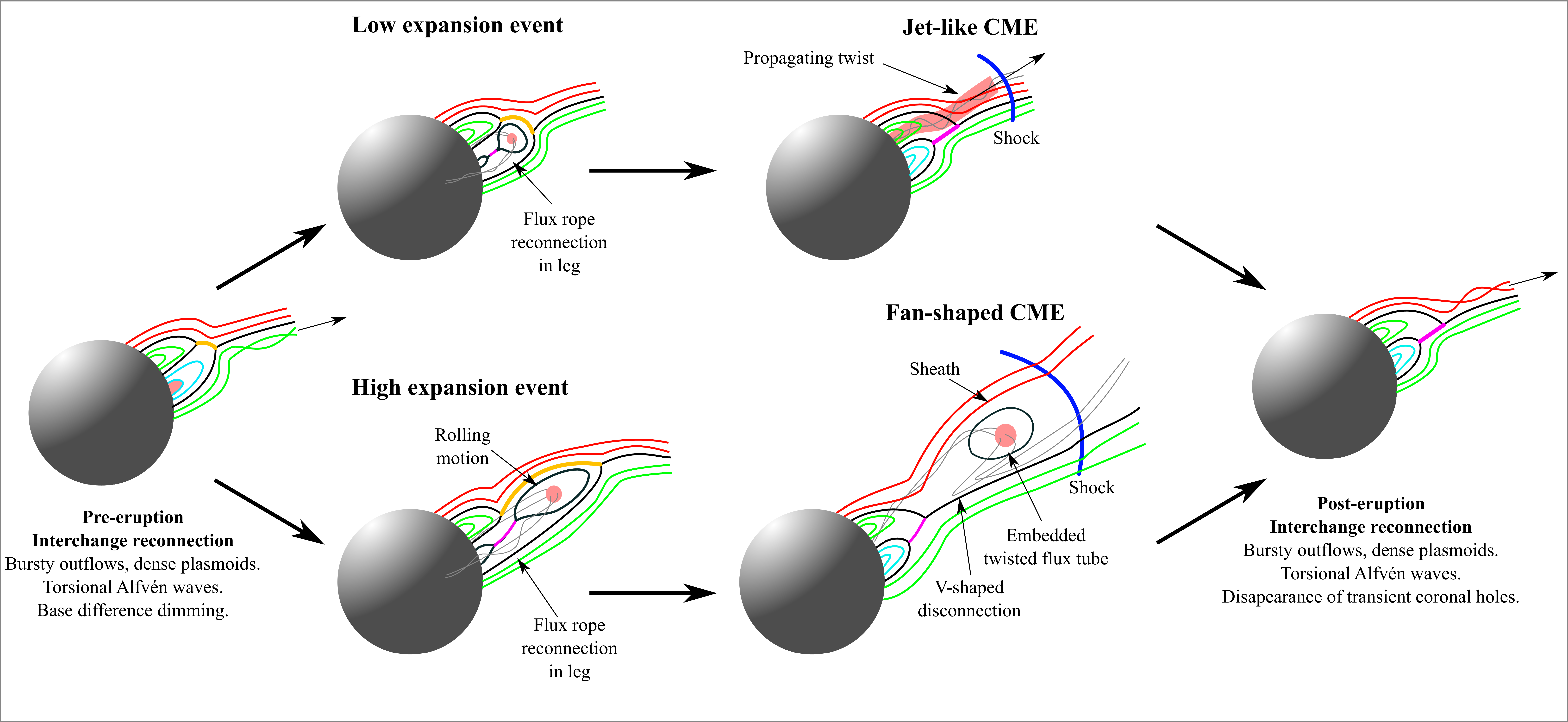}
    \caption{Schematic summary of pseudostreamer CMEs combining our latest results with previous observational and simulation studies. {Top row: low-expansion, jet-like CME events. Bottom row: high-expansion, fan-shaped CME events.}}
    \label{fig:schem}
\end{figure*}

\section{Discussion}
\label{sec:discussion}

\subsection{Implications for Theory}
To better understand the role played by the ideal flux-rope evolution in the pseudostreamer CME, we revisited findings from previous coronal-hole jet studies. For the vertical jet simulation reported in \citet{Wyper2018}, we calculated the average twist $\langle T_w \rangle$ and local decay index $n$ of the flux rope in the same manner as described above. The results are that  $\langle T_w \rangle \approx 2.2$ and $n\approx 2.0$ around the time that the vertical jet was launched. Both values are high enough to imply linear instability, and are comparable to those obtained for the CME. The breakout reconnection already had become self-sustaining for the jet, just as it had for the CME. In both cases, the upward flux-rope motion was slow, not explosive, and clearly was set by the rate at which breakout reconnection removed the strapping field above. The key signature associated with the transition to strong upward acceleration of the flux ropes was the onset of fast flare reconnection in the corona below them.

A contrast to these behaviors was found in a related study of active-region periphery jets, in which the ambient magnetic field is highly inclined from the vertical \citep{Wyper2019}. In such configurations, the null point resides in the low corona off to the side of the filament-channel flux rope, rather than in the high corona above it. We found in that case that the breakout feedback was inhibited; the flux rope rose more or less vertically within the pseudostreamer dome rather than pressing its strapping field horizontally against the breakout current layer. Lacking effective feedback between the flux-rope rise and the breakout reconnection, eventually the flux rope suffered a classic kink instability. Its internal twist converted to writhe of its axis, distorting the flux rope into an inverse-$\gamma$ shape as it kinked, and its apex tilted toward the null point and accelerated to and, later, through the breakout current layer there \citep[see Figure 9 in][]{Wyper2019}. The internal twist peaked at $T_w \approx 3$ and averaged $\langle T_w \rangle \approx 1.5$ \citep[see Figure 10 and text in][]{Wyper2019}, values that are similar to those for the vertical jet and the CME. The rapid rise of the newly kinked flux rope in this third case was accompanied, rather than preceded, by the onset of fast flare reconnection below it.

The absence of the expected signatures of kink-instability onset, which are so clear in the active-region-periphery (ARP) jet, from the vertical jet and the CME suggest that the instability plays no critical role in either case. The evidence is less clear-cut for the lack of a critical role for torus instability, but as noted, the critical index $n$ measured at our flux ropes is well above the typically cited threshold $n = 1.5$ \citep{Kliem2006}. The same is true for the twist parameter $T_w$ and its profile-dependent threshold $T_w \approx 1.5$ for the kink instability. What do these facts imply?

One possibility is that the analytically derived thresholds are too small, and the actual thresholds are significantly higher. This is plausible, given the special symmetry of the equilibrium configurations and the simplifying assumptions required by the analyses. Our configurations possess no special symmetries whatsoever, and the assuredly stabilizing effects of line-tying the overlying strapping fields must be taken into account but greatly complicate the analyses. The main argument against this explanation is the occurrence of classic kink signatures in the ARP jet, although it certainly is possible that only this particular example among our three cases actually reaches its true instability threshold.

A second possibility is that either instability has, or both have, in fact, reached the true threshold for onset; but the evolving system has adjusted to attain a quasi-static state in which the mode(s) saturated. The overlying strapping fields have more freedom to expand upward and accommodate the strengthening flux rope in the vertical jet and the CME, with the null point high above, than in the ARP jet, with the null point low and off to the side. The last case may simply constrain the flux rope so much more effectively that it becomes strongly unstable and kinks violently, unlike the other two cases.

Indirect support for the above explanations is provided by the translationally symmetric simulation of pseudostreamer CMEs by \citet{Lynch2013}. Their eruptions were driven by sheared arcades: flux-rope instabilities played no role because there were \emph{no flux ropes} in the system prior to eruption. Due to the special symmetry, the CME flux ropes formed only upon onset of the flare reconnection, and they were untethered to the Sun at creation. The ideal expansion of the increasingly energized sheared arcade field eventually induced breakout reconnection of the strapping fields at the null point above. The expansion turned explosive when fast flare reconnection switched on below the rising arcade, rapidly accelerating both the ideal upward motion and the breakout reconnection at the apex of the pseudostreamer, and forming the untethered CME flux rope in the process.

A fully definitive resolution of the role of ideal instability in these simulated eruptions cannot be achieved with Eulerian MHD models, such as ARMS, alone. All such computational models have irreducible amounts of numerical diffusion in them to stabilize their solutions and make them monotone. The consequence is that it is impossible to eliminate all nonideal evolution from their calculations, including magnetic reconnection. A purely ideal model is needed to simulate these configurations, and others, to firmly determine whether ideal instability is essential or inconsequential to the initiation of coronal jets and CMEs. The Lagrangian Field-Line Universal relaXer \citep[FLUX;][]{deforest2007,lowder2024} would be a possible tool to apply to such studies \citep[e.g.,][]{Rachmeler2010}.

\subsection{Implications for Observations}
The present simulation explains many observational features of fan-shaped pseudostreamer CMEs. Prior to eruption, the model predicts that both a faint jet and a base-difference dimming should be produced along the open spine as breakout reconnection launches closed-field plasma into the heliosphere, as has been described previously \citep{Kumar2021,Wyper2021}. The model also reproduces the bursty outflows and dense plasmoid signatures often observed in the breakout and post-eruption flare current sheets \citep{Kumar2019a,Kumar2021,Kumar2023}.  The model further predicts a rolling motion of the erupting flux rope in the low corona. The flux rope is deflected toward the lower field strength at the null point, as has been noted in previous studies \citep{Panasenco2011,Lynch2013,Sahade2022}, but also rotates due to the flare reconnection jet becoming oriented along the side of the flux rope, Figure \ref{fig:panels}(c). Such rolling motions are common in the early stages of fan-shaped pseudostreamer CMEs \citep{Wang2018,Kumar2021}. 

The simulation produces V-shaped features that may explain those noted by \citet{Wang2018}. First, V-shaped retracting field lines are formed on the underside of the flux rope in the early stages of the eruption; e.g., Figure \ref{fig:panels}(g). Second, much larger V-shaped field lines are formed when the flux rope disconnects; e.g., Figure \ref{fig:disc}(d). The disconnection process also produces retracting high-lying cusp structures, which \citet{Wang2018} concluded were evidence of interchange reconnection. The simulation reveals that the fan-shaped CME is an open-field magnetic structure, with an embedded twisted flux tube that is the remnant of the original coronal flux rope. Embedded bubble-like structures are sometimes observed in these events \citep[cf.\ Fig.\ 2, third row, in][]{Wang2015}. Furthermore, \citet{Wang2018} noted that the CME exhibited a ``twisting'' motion concurrent with the formation of the high-lying cusp structures. Our simulation shows that this twisting is likely the whip-like motion of the disconnected flux-rope and sheath field lines. This differs from the spire rotation along a fixed spine seen in jet-like CMEs and coronal jets associated with filament channel eruptions. Finally, the simulation predicts that a shock is created and propagates out ahead of the CME body. Such shocks have been observed ahead of blowout jets and pseudostreamer CMEs in white-light images \citep[e.g.,][]{Vourlidas2003,Miao2018,Kumar2021}. In a follow-up study we will explore in greater detail the synthetic white-light signatures of this simulation. 

Our simulation also offers insight into how high-energy particles could be released into the heliosphere in these events. The opening of the flux rope here is similar to the scenario established by \citet{Masson2013,Masson2019} for the release of high-energy particles. In their case, the null-point topology was fully closed beneath a helmet streamer initially, then dynamically opened during the eruption. In our case, the null-point topology is surrounded by open field from the outset. In both cases, however, high-energy particles accelerated by closed/closed flare reconnection in the early stages of the eruption will be trapped within the flux rope. Some will mirror back and forth between the two footpoints. When the disconnection occurs, these particles will be promptly released out along the newly opened field. Some particles may be accelerated further by the intense interchange reconnection associated with the disconnection. Such particle bursts should be detectable in situ by missions such as Solar Orbiter or Parker Solar Probe. They may also be associated type-III radio bursts \cite[e.g.,][]{kumar2017,Chen2018} in the same way that many coronal jets are.

The trailing tail of the CME is dominated by torsional Alfv\'{e}nic waves and denser field-aligned flows associated with the interchange reconnection that enables the flux disconnection. The Alfv\'{e}nic waves might steepen to form switchbacks \citep{Squire2020,Wyper2021} that could be detected as a switchback patch by Parker Solar Probe. Furthermore, the enhanced density in the field-aligned flows should be observable with high-cadence white-light coronagraphs on missions such as Solar Orbiter. Production of these waves and flows will continue for several hours after the CME has concluded. Similar post-eruption interchange reconnection, but on the much smaller scales of coronal jets, leads to the formation of transient plumes \citep{Raouafi2014}. This correspondence also highlights the similarities between jets and pseudostreamer CMEs.

\section{Summary}
We have presented an analysis of a simulated broad, fan-shaped pseudostreamer CME. Based on our findings, in Figure \ref{fig:schem} we summarise the key features of these eruptions and how they compare with narrow, jet-like pseudostreamer CMEs. Both types are ultimately constrained by the adjacent open field and have a discernible element of rotation. In jet-like CMEs, the rotation manifests in the propagation along the spire of twist from the flux rope in the form of helical outflows, whereas in fan-shaped CMEs it manifests as a whip-like motion of the embedded twisted flux tube (the remnant of the original coronal flux rope) and its sheath field lines. Both CME types are comprised of open field lines following the disconnection of one end of the flux rope due to interchange reconnection. In addition, both are preceded and followed by bursty interchange reconnection that launches torsional Alfv\'{e}nic waves and episodic field-aligned dense outflows. 

The key difference between the two types of pseudostreamer CMEs is the greater expansion of the erupting flux rope in broad, fan-shaped versus narrow, jet-like eruptions. At larger scales where the spherically expanding geometry is important, the field strength falls off with height allowing for greater transverse and vertical expansion of the developing flux rope. This enables the eruption to more easily push aside the ambient background field so that more of the flux rope survives intact its ascent into the high corona, forming a fan-shaped CME. At smaller scales and in a straighter, more uniform ambient field, the flux rope is more highly constrained by and interacts more strongly with the background field. More of the flux rope is consumed by reconnection as it breaches the null point, injecting its twist onto the surrounding open field and forming a collimated, narrow CME.

Despite their differing CME morphologies, the magnetic connectivity evolution is the same in both event types: it consists of the eruption and disconnection of a flux rope from beneath the pseudostreamer topology. Therefore, our model shows that fan-shaped pseudostreamer CMEs are simply the extreme end of a continuum of eruptive events which includes jet-like pseudostreamer CMEs and minifilament coronal jets. Moreover, these eruptive events should be considered as a separate class of CMEs from the bubble-like CMEs that originate beneath helmet streamers. In those events, the connection of the flux rope at both ends to the solar surface is maintained to much greater distances from the Sun owing to the magnetic polarity reversal across the heliospheric current sheet at the top of the helmet streamer. 

This work highlights the need for coordinated simulation and observational studies of pseudostreamer CMEs. In a follow-up paper (Lynch et al. 2024) we will present the expected remote and in-situ observational signatures of this simulation as a guide for interpreting the latest observations from Solar Orbiter and Parker Solar Probe. We also plan to conduct future simulations to identify the conditions that govern the transition from narrow, jet-like to broad, fan-shaped CMEs in these pseudostreamer events.

\section*{} 
We thank Judy Karpen, Angelos Vourlidas, Mariana Cec\'{e}re, and Etienne Pariat for useful discussions. PFW was supported by an STFC (UK) consortium grant ST/W00108X/1 and a Leverhulme Trust Research Project grant. PFW \& BJL were supported by NSF grant NSF AGS 2147399. BJL acknowledges support from the NASA XRP and LWS programs. CRD was supported by a NASA H-ISFM grant to Goddard Space Flight Center and the NASA XRP program. SKA was supported by a LWS grant to U Michigan. The computations were sponsored by allocations on Discover at NASA's Center for Climate Simulation and on the DiRAC Data Analytic system at the University of Cambridge, operated by the University of Cambridge High Performance Computing Service on behalf of the STFC DiRAC HPC Facility (www.dirac.ac.uk) and funded by BIS National E-infrastructure capital grant (ST/K001590/1), STFC capital grants ST/H008861/1 and ST/H00887X/1, and STFC DiRAC Operations grant ST/K00333X/1. DiRAC is part of the National E-Infrastructure. The supporting ARMS dataset is published in Durham University's Collection open data repository.\dataset[DOI: 10.15128/r1fb494847x]{http://doi.org/10.15128/r1fb494847x}








\end{document}